\documentstyle[psfig,11pt,fullpage]{article}
\begin{document}

\begin{center}
\thispagestyle{empty}
{\Large{\bf Spinning Black Holes in $(2+1)$-dimensional\\
String and Dilaton Gravity}}\\
\vspace{1.5cm}
Kevin C.K. Chan\footnote
{\normalsize Email address: kckchan@avatar.uwaterloo.ca}
and {Robert B. Mann}\footnote
{\normalsize Email address: mann@avatar.uwaterloo.ca}\\
Department of Physics, University of Waterloo,\\
 Waterloo, Ontario, Canada N2L 3G1.\\
\end{center}
\vspace{0.5cm}
\centerline{ABSTRACT}
\bigskip

We present a new class of spinning black hole solutions
in $(2+1)$-dimensional general relativity minimally coupled to a dilaton with
potential
$e^{b\phi}\Lambda$. When $b=4$, the corresponding
spinning black hole is a solution of low energy $(2+1)$-dimensional string
gravity. Apart from the limiting case of the $BTZ$ black hole,
these spinning black holes have no inner horizon
and a curvature singularity only at the origin.
We compute the mass and angular momentum
parameters of the solutions at spatial infinity, as well as their
temperature and entropy.

\vspace{0.2 cm}



\vspace{0.5cm}

\section{Introduction}

Lower dimensional gravity is a source of fascination
for theorists, primarily because of the potential insights
into quantum gravity that it offers. The lower-dimensional setting affords
a significant amount of technical simplification of the gravitational field
equations, bringing into sharper focus conceptual issues that are often
obscured in the more complicated $(3+1)$-dimensional case.
In particular, there has been great interest in recent years in
$(2+1)$-dimensional general relativity ever since the discovery of the $BTZ$
black hole
solution \cite{btz}. This $BTZ$ black hole
has so far attracted much interest in its classical, thermodynamic,
statistical and quantum properties (see \cite{carlip} for a review).
It is also a solution to the low energy
string theory \cite{welch}.

It is interesting and important to investigate how the
properties of black holes are modified when a dilaton is
present. The first such modification of the non-spinning charged
$BTZ$ black hole was investigated in \cite{chaman} with
an exponential potential and an asymptotically non-constant dilaton.
In addition, by considering several asymptotically constant dilatons and
different
forms of potential functions, some interesting modifications
of the $BTZ$ black hole can be found \cite{chan}.
The action considered in \cite{chaman} is given by
\begin{equation}
S = \int  d^3x {\sqrt{-g}}
({\cal R}-4(\nabla\phi)^2+2e^{b\phi}\Lambda-e^{-4a\phi}F^2), \label{act}
\end{equation}
with arbitrary couplings $\Lambda$, $a$ and $b$, where ${\cal R}$ is the
Ricci scalar, $\phi$ is the dilaton field and $F^2$ is the usual Maxwell
contribution. The constants $a$ and $b$ govern the coupling
of $\phi$ to $F^2$ and $\Lambda$ respectively.
For a constant $\phi$, (\ref{act}) admits the usual $BTZ$ case, where
$\Lambda$ is the cosmological constant [note that
in the presence of a non-trivial dilaton,
the space does not behave as either de Sitter ($\Lambda<0$) or anti-de
Sitter space ($\Lambda>0$)].
When $a=1$ and $b=4$, the action (\ref{act}) is that of low energy
string theory in terms of the Einstein metric.
The corresponding action in terms of the string metric
can be obtained by performing the conformal transformation \cite{shar}
\begin{equation}
g^{S}_{\mu\nu} = e^{4\phi}g^{E}_{\mu\nu}, \label{conformal}
\end{equation}
where $S$ and $E$ denote the string and Einstein metrics respectively.
The corresponding string action is
\begin{equation}
S = \int  d^3x\sqrt{-g^S}e^{-2\phi}({\cal R}[g^S]
+4(\nabla\phi)^2-F^2+2\Lambda). \label{sact}
\end{equation}
By taking the product of the $(1+1)$-dimensional string-theoretic black hole of
Mandal,
Sengupta and Wadia \cite{witten} with ${\bf{S^1}}$, a
static charged black hole solution to the field equations of
string action (\ref{sact}) with $\phi=-{1\over 4}ln(r)$
may be obtained (hereafter referred as $2+1$ $MSW$ black hole)
-- if a product with ${\bf {R}}$ is taken instead,
one gets a black string \cite{gary}.

The black hole solutions obtained in
\cite{chaman} belong to (\ref{act}) with the choice
$\phi\propto ln(r)$ and they
contain the non-spinning $BTZ$ and
$2+1$ $MSW$ black holes as special cases.
In fact, (\ref{act}) contains many other dilaton solutions (without black
holes)
which were previously investigated by a number of authors (see references
in~\cite{chaman}). Note that (\ref{act}) can be viewed as general
relativity with an unusual matter Lagrangian. For example, one can
easily see that the local energy density in the perfect fluid form is
negative when $\phi=constant$ and $\Lambda>0$.

In this paper, by adopting the same choice of dilaton
as in the static black holes in \cite{chaman},
we obtain a family of spinning black holes
as solutions to the field equations which follow from
action (\ref{act});
in particular, the $b=4$ spinning black hole belongs to the
string theory (\ref{sact})\ with the conformal transformation
(\ref{conformal}).
For simplicity, we ignore the Maxwell fields in this paper;
the possibilities of including Maxwell fields will be investigated in a future
work.
The solutions we derive in this paper are not vacuum solutions
in a strict sense since the action (\ref{act}) contains a static
``dilaton fluid'' whose energy momentum tensor is nowhere vanishing.
Here we adopt special dilaton fluid models in which the metric
approaches neither the Minkowski nor anti-de Sitter metrics at
spatial infinity similar to the previous black holes obtained
in \cite{chaman} and \cite{chm}. It is worthwhile mentioning that in recent
a naive argument \cite{chan} demonstrates that there exist no asymptotically
flat positive mass black hole solutions in $2+1$ dimensions.
Although the dilaton field diverges for large $r$
($\phi\propto \ln r$), the
(quasilocal) mass and angular momentum are finite for all values of $r$
outside the event horizon in every black hole solution.
The Ricci and Kretschmann scalars are everywhere finite
except at $r=0$ but those singularities are hidden by the event
horizon. It can also be checked that all terms in the action (\ref{act})
are finite for $\infty>r > r_h$ (where $r_h$ denotes the event horizon),
and vanish as $r\rightarrow\infty$ in all the black hole solutions we obtain.
As a consequence we believe that our spinning black hole solutions are of some
interest.

The organization of this paper is as follows. In section 2 we quickly review
the static and circularly symmetric black hole solutions obtained
in \cite{chaman}. The quasilocal formalism for calculating energy, mass and
angular momentum will also be briefly presented in this section.
In section $3$, we present the spinning black hole solutions.
Various properties of the black hole metrics are discussed and their
temperatures are also computed. We will see how a non-trivial dilaton
modifies the spinning BTZ black hole. We summarize our work in a concluding
section.

Our conventions are as in Wald \cite{wald}; we set the gravitational
coupling constant $G$ equal to unity. The signature of the metric is $(-++)$.

\section{Review of Static Dilaton Black Holes and Quasilocal Formalism}

We now consider the $2+1$-dimensional action (\ref{act}) and review its
uncharged
dilaton black hole solutions. We refer readers to \cite{chaman} for the charged
cases.
Varying (\ref{act})
with respect to the metric and dilaton fields, respectively,
yields (after some manipulation)
\begin{equation}
{\cal R}_{\mu\nu} =  4\nabla_{\mu}\phi\nabla_{\nu}\phi -
2g_{\mu\nu}e^{b\phi}\Lambda, \label{eoma}
\end{equation}
\begin{equation}
4\nabla^2\phi =  -be^{b\phi}\Lambda. \label{eomb}
\end{equation}
A family of static, circularly symmetric black hole
solutions to these field equations was obtained in \cite{chaman}.
They are given by
\begin{equation}
\phi = k ln(r),\label{dilaton}
\end{equation}
\begin{equation}
ds^2 = -U(r)dt^2+\beta^2{dr^2\over U(r)} + \beta^2 r^Nd\theta^2, \label{met}
\end{equation}
with
\begin{equation}
U(r) = \left[-{2M\over N}r^{1-{N\over 2}} + {8\Lambda\beta^{2}r^N\over
(3N-2)N}\right],\label{sol}
\end{equation}
and where
\begin{equation}
k = \pm{1\over 4}\sqrt{N(2-N)}, \label{const}
\end{equation}
and
\begin{equation}
bk = N-2. \label{coupling}
\end{equation}
$M$ is the quasilocal mass identified at spatial infinity.
Positive mass ($M>0$) black hole solutions only
exist for $2\geq N>{2\over 3}$ and $\Lambda>0$.
It is obvious that as $r\rightarrow\infty$,
$\phi\rightarrow\infty$ too. However, the kinetic term
and the potential term in the action (\ref{act})
all vanish in that limit when $2\geq N>0$.
When $N=1$, (\ref{sol}) reduces to the $2+1$ the uncharged $MSW$ black hole;
if $N=2$, it becomes the uncharged $BTZ$ black hole which is also a
solution to string theory. In general, (\ref{met}) and (\ref{sol}) are neither
asymptotically flat
nor anti-de Sitter. Note that the radial co-ordinate $r$ is chosen to be
dimensionless and $\beta$ is a length scale with dimensions of length
$[L]$.

We now present the formulas for calculating the quasilocal mass, angular
momentum and energy of a stationary and axisymmetric
solution in an asymptotically non-flat spacetime; for a detailed derivation
of these, see \cite{broyor}. For a ($2+1$)-dimensional stationary
and axisymmetric spacetime, the metric can be written as \cite{kawa}
\begin{equation}
ds^2 = -L^2dt^2 + f^{-2}dR^2 + R^2(N^{\theta}dt + d\theta)^2, \label{sa}
\end{equation}
where $L(R)$, $f(R)$ and $N^{\theta}(R)$ are functions of $R$ only --
under a co-ordinate transformation $R=R(r)$, they can be functions of a new
radial variable $r$. $L(R)$ is called the lapse function and $N^{\theta}(R)$
is called the angular shift.
Using (\ref{sa}) and the formalism adopted in \cite{broyor}, the quasilocal
energy
$E(R)$ at a radial boundary $R$ can be shown to be
\begin{equation}
E = 2(f_o(R)-f(R)).\label{energy}
\end{equation}
Here $f^2_o(R)=g^{RR}_o(R)$ is a background metric function which determines
the zero of the energy. The background metric can be obtained simply
by setting constants of integration of a particular solution to some special
value that then specifies the reference spacetime. The quasilocal mass
$m(R)$ at a $R$ is given by
\begin{equation}
m = L(R)E - j(R) N^{\theta}(R),\label{mass}
\end{equation}
where $j(R)$ is the quasilocal angular momentum given by
\begin{equation}
j = {f(R)N^{\theta '}(R)R^3\over L(R)}, \label{ang}
\end{equation}
where the prime denotes the ordinary derivative with respect
to $R$. As $R\rightarrow\infty$, the analogous $ADM$ mass and angular momentum
are defined to be $M\equiv m(\infty)$ and $J\equiv j(\infty)$ respectively.
Note that $E(R)$, not $m(R)$, is the thermodynamic internal energy
\cite{broyor}.
This distinction between mass and energy is important for spacetimes which
are not asymptotically flat, since in those cases the magnitude of the
timelike killing vector field diverges as it approaches spatial infinity.
The thermodynamic temperature $T(R)$ at a given value of $R$ is defined as
\begin{equation}
T = {\partial E\over\partial S},\label{temp}
\end{equation}
where $S$ is the entropy of the black hole. General arguments
similar to
those in $3+1$ dimensions (see, {\sl e.g.} \cite{visser}) show that
\begin{equation}
S = 4\pi R_+. \label{entropy}
\end{equation}
Thus using (\ref{energy}), (\ref{temp}) and (\ref{entropy}), the temperature
can be calculated easily.
The analogous Hawking temperature is obtained by taking
the spatial infinity limit.
It can be shown that the temperature calculated using (\ref{temp}) is the usual
surface
gravity divided by a red-shift factor. For asymptotically flat spacetimes,
the red-shift factor tends to unity at spatial infinity.

\section{Spinning Black Hole Solutions}

We now present a family of spinning black hole solutions characterized by
the mass $M$, angular momentum $J$ and dilaton coupling $N$ parameters.
We use the same dilaton (\ref{dilaton}) and relationships among $b$, $K$ and
$N$ in (\ref{const}) and (\ref{coupling}). In terms of the same radial
co-ordinate
$r$ as in the static case (\ref{met}) and (\ref{sol}), the spinning solutions
are
\begin{eqnarray}
ds^2 & = & -\left({8\Lambda\beta^2r^N\over (3N-2)N} + Ar^{1-{N\over
2}}\right)dt^2
+{\beta^2dr^2\over\left[{8\Lambda\beta^2r^N\over (3N-2)N}
+\left(A-{2\Lambda\omega^2\over (3N-2)NA}\right)r^{1-{N\over 2}}\right]}
\nonumber \\ & &
 -\omega r^{1-{N\over 2}}dtd\theta
+\left(\beta^2r^N-{\omega^2\over 4A}r^{1-{N\over 2}}\right)d\theta^2,
\label{spinsola}
\end{eqnarray}
\begin{equation}
M =  {N\over 2}\left[{2\Lambda\omega^2\over (3N-2)NA}\left({4\over
N}-3\right)-A\right], \label{spinsolb}
\end{equation}
\begin{equation}
J =  {3N-2\over 4}\omega. \label{spinsolc}
\end{equation}
Using (\ref{sa}) and the $g^{rr}$ in (\ref{spinsola}), the lapse, $f(r)$,
angular
shift and radial functions are
given by
\begin{equation}
L^2 = {g^{rr}\beta^4\over \left(\beta^2-{\omega^2\over 4A}r^{1-{3N\over
2}}\right)}, \label{lapse}
\end{equation}
\begin{equation}
f^2 = g^{RR} = g^{rr}\left({dR\over dr}\right)^2, \label{func}
\end{equation}
\begin{equation}
N^{\theta} = -{\omega r^{1-N/2}\over 2R^2}, \label{shift}
\end{equation}
and
\begin{equation}
R^2 = g_{\theta\theta} = \left(\beta^2r^N-{\omega^2\over 4A}r^{1-{N\over
2}}\right). \label{radial}
\end{equation}
$A$ and $\omega$ are two integration constants which are related to the mass
$M$
and angular momentum $J$ respectively in (\ref{spinsolb}) and (\ref{spinsolc}).
It is easy to check (see below) that black holes exist if
$\Lambda>0$ and $2\geq N >{2\over 3}$.
Except for the $N=1$ and $N=2$ cases, we cannot explicitly express $r$ in terms
of $R$. As a result, all metric functions in the above are not
explicit functions of $R$. However, when $r\rightarrow\infty$,
by using binomial expansion and reversion of the series,
an approximation is given by $r^{-1}=\left({\beta\over R}\right)^{{2\over
N}}-{\omega^2\over 4A\beta^2 N}\left({\beta\over R}\right)^{3}+\cdot\cdot\cdot$
for the range $2\geq N >{2\over 3}$.
The radial transformation
(\ref{radial}) is based on the assumption that $g_{\theta\theta}$ part in
(\ref{spinsola})
must always be greater than zero. However, if $g_{\theta\theta}<0$ is allowed,
one must have $g_{\theta\theta}d\theta^2=-R^2d\theta^2$. Hence $\theta$ is a
timelike co-ordinate, and because it is periodic, it corresponds to a region
with closed timelike curves. To avoid this,
$A$ must be chosen to be negative in (\ref{spinsolb}) so that
$g_{\theta\theta}>0$.
Thus $2A=-{2M\over N}-\sqrt{{4M^2\over N^2}+\left({4\over N}-3\right)
{8\Lambda\omega^2\over (3N-2)N}}$.
$M$ is identified using (\ref{mass}) and the expansion of ${1\over r}$ above
when the
background metric $f_o=f(A=\omega=0)$ is chosen. It is taken to be positive.
$J$ is identified at spatial infinity ($r\rightarrow\infty$) as well.
This choice of background is consistent with the fact that
$M$ is the mass parameter at spatial infinity
identified by Brown, Creighton and Mann \cite{broyor} for the $BTZ$ black hole
when $N=2$. (\ref{spinsola}) reduces to the static solution (\ref{sol}) when
$J=0$.
We can see that the presence of $J$ modifies the static solutions by
introducing
an extra term in $g_{\theta\theta}$ and by causing the
coefficients of the $r^{1-{N\over 2}}$ term in $g_{tt}$ and
$g_{rr}$ to become unequal. When $N=2$, using (\ref{spinsola}),
(\ref{spinsolb}) and (\ref{spinsolc}), and the radial co-ordinate
transformation
(\ref{radial}), it is easy to show that
\begin{equation}
ds^2 = -(\Lambda R^2-M)dt^2
+\left(\Lambda R^2-M+{J^2\over 4R^2}\right)^{-1}dR^2
-Jdtd\theta+R^2d\theta, \label{btzmet}
\end{equation}
which is exactly the spinning $BTZ$ black hole with mass $M$ and angular
momentum
$J$.

One can locate the event horizon(s) from the vanishing of
the lapse in (\ref{lapse}).
It can be checked that (\ref{spinsola}) admits an event horizon if the
following
conditions are satisfied:  $M>0$, $\Lambda>0$ and $2\geq N>{2\over 3}$.
In terms of $r$, (\ref{lapse}) vanishes at $r_+$:
\begin{equation}
\left({4\over N}-3\right)
{4\Lambda\beta^2\over (3N-2)N}r_+^{{3N\over 2}-1}=
{M\over N}\left({2\over N}-1\right)+\sqrt{{4M^2\over N^2}
+\left({4\over N}-3\right){8\Lambda\omega^2\over (3N-2)N}}
\left({1\over N}-1\right) \label{outer}
\end{equation}
for $N\neq {4\over 3}$. If $N={4\over 3}$, then $r_+$ is given by
\begin{equation}
3\Lambda\beta^2 r_+ = {3M\over 2} - {\Lambda\omega^2\over 2M}. \label{outera}
\end{equation}
Using (\ref{spinsolb}), (\ref{spinsolc}), (\ref{radial}) and (\ref{outer}) with
$N=2$, it can be
shown that $R_+^2={M+\sqrt{M^2-\Lambda J^2}\over 2\Lambda}$ which is the
location of the outer event horizon of the $BTZ$ black hole in terms of the $R$
co-ordinate. The lapse function also vanishes at the ``inner horizon''
\begin{equation}
r_- = 0. \label{inner}
\end{equation}
$r_-=0$ corresponds to a zero circumference (see (\ref{radial})),
and thus it is irrelevant.  However, when $N=2$,
$r_-=0$ implies $R_-^2=-{J^2\over 4A}$ in (\ref{radial}) and the circumference
of it
is not zero. Using (\ref{radial}), it can further be shown that
$R_-^2={M-\sqrt{M^2-\Lambda J^2}\over 2\Lambda}$
which is exactly the expression for the location of the inner horizon of the
spinning $BTZ$ black hole. Thus except for the $BTZ$ ($N=2$) case,
the family of spinning black holes (\ref{spinsola}) has no inner horizon
even if $J\neq 0$.
In addition, except for the $BTZ$ ($N=2$) case, where there are no divergences
in
curvature invariants, it is lengthly but straightforward
to show that for $2\geq N>{2\over 3}$,
the scalar and Kretschmann scalars only diverge at $r=r_{-}=0$ ($R=0$).
Thus as long as $r_+>0$, the solutions do not have naked singularities.
The extremal limit between $M$ and $\omega$ ($J$) can be defined as the
vanishing of
$r_+$ in (\ref{outer}) and (\ref{outera}) for non-vanishing $M$ and $J$.
In (\ref{outer}), there is no extremal limit for
$1\geq N>{2\over 3}$. $r_+$ is always real and greater than zero.
Previous examples of black holes without extremal limit between mass and charge
can be found in \cite{chm}. For the range $2\geq N>1$, $r_+$ is real and
greater than
zero in (\ref{outer})  if
\begin{equation}
M^2\geq (N-1)^2{8\Lambda\omega^2\over (3N-2)N} \label{extreme}
\end{equation}
where the equality sign corresponds to the extremal limit.
When $N=2$, $M^2=\Lambda J^2$ as expected.
In (\ref{outera}), the extremal limit simply is
\begin{equation}
3M^2 = \Lambda\omega^2. \label{extremea}
\end{equation}
Solutions (\ref{spinsola})-(\ref{spinsolc}) contain three particular
interesting cases. First,
$N=2$ is the $BTZ$ case as
we have just shown. Second, the $N={4\over 3}$ solution
is conformally related to the the black hole solution previously derived
by Lemos in \cite{lemos}. It was shown in \cite{lemos} that his
black hole solution can be translated to the $3+1$ spacetime
as a cylindrical solution to a $(3+1)$-dimensional
dilaton gravity. The third case is the $N=1$ solution which
corresponds to the low energy $2+1$ string theory. In terms of the Einstein
metric, the solution reads
\begin{eqnarray}
ds^2 & = & -\left(8\Lambda\beta^2r - {2M+\sqrt{4M^2+8\Lambda\omega^2}
\over 2}\sqrt{r}\right)dt^2
+{\beta^2dr^2\over\left[8\Lambda\beta^2r-2M\sqrt{r}\right]} \nonumber\\ & &
 -\omega \sqrt{r}dtd\theta
+\left(\beta^2r+{-2M+\sqrt{4M^2+8\Lambda\omega^2}\over
16\Lambda}\sqrt{r}\right)d\theta^2,  \label{smetrica}
\end{eqnarray}
\begin{equation}
J =  {1\over 4}\omega.  \label{smetricb}
\end{equation}
This $N=1$ spinning black hole is a modification of the $2+1$
$MSW$ static black hole. Using the conformal transfomation
(\ref{conformal}), one can express the $N=1$ black hole solution in terms of
the
string metric. It is the second example of a non-trivial
black hole solution in $2+1$ string theory (the first one is the $BTZ$ case).
Note that in terms of the string action (\ref{sact}),
one can see that the string coupling $e^{2\phi}={1\over \sqrt{r}}$ vanishes as
$r\rightarrow\infty$. Due to the presence of a non-vanishing $J$, the
dimensional
reduction of the $N=1$ string solution is no longer a solution to $1+1$ string
theory, since one expects that $J$ will introduce an extra potential term in
the original $1+1$ string action.

We can deduce the causal structures of the spinning dilaton black holes (except
for the $BTZ$ case) as follows.
The casual structures of the static uncharged black holes in (\ref{sol}) were
drawn
in \cite{chaman}. They all have a spacelike
singularity at $r=0$. Furthermore they either have
a timelike or null-like null infinity, depending on the value of $N$. In the
former case the causal structure is the same
as the non-spinning $BTZ$ one, while in the latter case it is the same as the
Schwarzscshild one. In the present spinning solutions  they all asymptotically
approach the static uncharged black hole solutions (\ref{sol}), and they have
no inner
horizons.
There is a spacelike curvature singularity at $r=0$. Thus one can deduce that
the spinning dilaton black holes would have similar causal structures either to
the non-spinning $BTZ$
or Schwarzschild case. We will not repeat the drawing here.

Finally, we briefly discuss the thermodynamic properties of the
spinning solutions.
An important thermodynamic quantity in a stationary black hole
is the temperature $T$ defined in (\ref{temp}).
Using the $r$ co-ordinate, one can show that (\ref{temp}) at a radial boundary
generally yields the surface gravity ($\kappa$) term
\begin{equation}
2\pi T=\kappa = {N\over 2R_+}\left({3N-2\over 4-3N}\right)
\left[{M\over N}\left({2\over N}-1\right)+\sqrt{{4M^2\over N^2}
+\left({4\over N}-3\right){8\Lambda\omega^2\over (3N-2)N}}
\left({1\over N}-1\right)\right] \label{surface}
\end{equation}
along with a redshift factor which may be computed as in \cite{broyor},
and $R_+$ is defined in (\ref{radial}) with $r_+$ is given by (\ref{outer}).
For (\ref{outera}), the temperature is
\begin{equation}
\kappa= {1\over 4R_+}\left(3M-{\Lambda\omega^2\over M}\right). \label{surfacea}
\end{equation}
It is easy to check that when $N=2$, $T$ in (\ref{surface})  reduces to the
$BTZ$
temperature obtained by Brown, Creighton and Mann in \cite{broyor}.
For the string case ($N=1$) with $J=0$, (\ref{surface}) is independent of the
mass
parameter
$M$. In the extremal cases (\ref{extreme}) and (\ref{extremea}), the
temperature vanishes in
(\ref{surface}) and (\ref{surfacea}).
The entropy can be trivially obtained using the entropy formula in
(\ref{entropy}).
Other thermodynamic quantities such as heat capacity and chemical potential can
be computed as in \cite{broyor}. We do not discuss them here.

\section{Conclusions}

We have obtained a family of asymptotically non-flat spinning
dilaton black hole solutions in $2+1$ with an exponential potential.
The family contains the spinning $BTZ$ metric as a special case.
In addition, one member of this class of black holes is a solution to
low energy string theory. All of these black hole spacetimes
have no inner horizons except the $BTZ$ case.
For the range $1\geq N>{2\over 3}$ of the parameter $N$,
the black holes have no extremal limit.
$J$ can take any finite value for a given $M$ without causing the event ho
rizon to disappear.

Although the static charged black hole solutions of (\ref{act}) exist
\cite{chaman},
at present we are unable to generalize our spinning solutions to charged cases.
This endeavour is complicated by the fact that when one adds Maxwell fields
to a spinning solution, both electic and magnetic fields must be present
\cite{kawa}.
As a result, the field equations are considerably more complicated to solve.
In order to simplify the differential equations involved,
a (anti) self dual condition on electric and magnetic fields
was recently imposed to get the spinning charged
BTZ black hole solution \cite{kawa}. However, by using (\ref{mass}) and
(\ref{ang}) in their
spinning solution,  it is easy to check that
the quasilocal mass and angular momentum, instead of being a
constant, diverges logarithmically at spatial infinity \cite{error}. Recently,
a class of static magnetic solutions to the Einstein-Maxwell equations
with $\Lambda>0$ was obtained, and it is horizonless and free of
curvature singularities \cite{hirsch}. It can be shown that the mass has a
logarithmic divergence at spatial infinity as well \cite{error}.
The logarithmic divergence is due to the fact that the electric potential is a
logarithmic function of $r$ in 2+1 dimensions.  One way of curing the
divergence is to add a topological
Chern-Simons term to the gauge action \cite{clement}. The resultant solution in
\cite{clement} is
horizonless, regular and asymptotic to the extremal $BTZ$ black hole.
Another way is to couple a dilaton to the Maxwell term in the action, similar
to (\ref{act}).
It was shown in \cite{chaman} that the static electrically charged black hole
solution has a constant electric term in $g_{tt}$ and $g_{rr}$ instead of a
logarithmic one.
Thus one of the roles of a dilaton in $2+1$ dimensions is to remove the
logarithmic divergence in the electric potential. It is an open question as to
whether one can get a charged version of the family of spinning string and
dilaton black holes obtained in this paper. We expect that the introduction of
electromagnetic fields
will yield  inner horizons to the present spinning dilaton solutions,
since the corresponding static
charged solutions in \cite{chaman} have inner horizons. One could then
attempt, say,  to derive exact mass inflation solutions to the charged
spinning solutions.
Early work on mass inflation for spinning $BTZ$ black hole
can be found in \cite{ccm}. We intend to relate further details elsewhere.

\bigskip
\centerline{\bf Acknowledgements}

This work was supported in part by the Natural Science and Engineering Research
Council of Canada. KCK Chan would like to thank Jim Chan for discussions.

\end{document}